\documentclass[10pt,conference,table]{IEEEtran} 
\usepackage[numbers]{natbib}
\usepackage{hyperref}
\usepackage{verbatim}
\usepackage{calc}
\usepackage{graphicx}
\usepackage{tcolorbox}
\usepackage{xcolor}
\usepackage{enumitem}
\IEEEoverridecommandlockouts

\title{Decentralized Self-Adaptive Systems: \\A Mapping Study}

\author{ 
\IEEEauthorblockN{Federico Quin} \IEEEauthorblockA{Department of Computer Science\\Katholieke Universiteit Leuven, Belgium
\\federico.quin@kuleuven.be} \and 
\IEEEauthorblockN{Danny Weyns} \IEEEauthorblockA{Katholieke Universiteit Leuven, Belgium
\\Linnaeus University, Sweden
\\danny.weyns@kuleuven.be} \and 
\IEEEauthorblockN{Omid Gheibi} \IEEEauthorblockA{Department of Computer Science\\Katholieke Universiteit Leuven, Belgium
\\omid.gheibi@kuleuven.be}}

\begin{document}

\maketitle

\begin{abstract}
With the increasing ubiquity and scale of self-adaptive systems, there is a growing need to decentralize the functionality that realizes self-adaptation. Our focus is on architecture-based self-adaptive systems where one or more functions for monitoring, analyzing, planning, and executing are realized by multiple components that coordinate with one another. While some earlier studies have shed light on existing work on the decentralization of self-adaptive systems, there is currently no clear overview of the state of the art in decentralization of self-adaptive systems. Yet, having a precise view on the state of the art in decentralized self-adaptive systems is crucial for researchers to understand existing solutions and drive future research efforts. To address this gap, we conducted a mapping study. The study focused on papers published at 24 important venues that publish research on self-adaptation. The study focused on the motivations for choosing a decentralized approach to realize self-adaptation, the adaptation functions that are decentralized, the realization of the coordination, and the open challenges in the area. 
\end{abstract}

\section{Introduction}

The upcoming generation of software-intensive systems will increasingly consist of loosely composed distributed entities, forming complex integrated and intelligent systems~\cite{ATZORI20102787,4519604,deLemos2013}. For example, the Internet-of-Things (IoT) exploits enhanced communication facilities and distributed intelligence to realize smart environments. The ability of such systems to provide the required level of guarantees for their quality goals throughout their lifetime will be of crucial importance~\cite{978-3-319-74183-3_2}. For instance, an IoT building security monitoring system needs to notify suspicious events in a timely and reliable way while  minimizing energy consumption, regardless of possible interference of communication links in the sensor network.

The challenges of such software-intensive systems have motivated the need for self-adaptation. Self-adaptation endows a system with the capability to adapt itself to dynamics in the environment, internal changes, and changes in its requirements in order to achieve particular quality goals~\cite{kephart2003vision,1186778.1186782,Cheng2009,weyns2020}. Central to the realization of self-adaptation is a feedback loop added to the system that monitors the managed system and its environment to collect knowledge about uncertainties at runtime that may have been difficult to predict at design time. This knowledge is then used to resolve uncertainties and adapt the managed system when needed to realize the quality goals. E.g., an IoT system that adapts communication paths to ensure reliability when the load in the network changes or parts of the network get affected by interference~\cite{10.1007}.

With the increasing ubiquity and scale of self-adaptive systems, there is a growing need to decentralize the functionality that realizes self-adaptation. Decentralization may for example come from the need to process data locally as the cost (e.g. in terms of required energy or bandwidth) to communicate all the data to one point may be too high. In this paper, we focus on self-adaptive systems that are based on a MAPE-based feedback loop~\cite{kephart2003vision}, which is a common approach to realize self-adaptation. Decentralization in such systems implies that one or more of the monitoring, analyzing, planning, and execution functions are realized by multiple components that coordinate with one another~\cite{weyns2012forms}. We focus on systems with MAPE-based feedback loops to draw coherent and consistent conclusions for this important field of self-adaptive systems. 

A substantial number of papers have studied decentralization in MAPE-based self-adaptive systems. In~\cite{1988008.1988037}, the authors extend MAPE-based feedback loops with support for intra-loop and inter-loop coordination. Another characteristic example is EUREMA~\cite{Vogel2014} that supports the specification of systems with multiple feedback loops and their coordinated execution. EUREMA is based on the notion of mega-model. Another example is DECIDE~\cite{Calinescu2015FSE}, a rigorous approach to decentralizing the control loops of distributed self-adaptive software. DECIDE relies on quantitative verification at runtime applied locally to meet  system-level quality-of-service requirements. A recent example is presented in~\cite{GEROSTATHOPOULOS201937} that introduces a runtime framework that can be implemented in a decentralized manner to dynamically modify, add and remove self-adaptation strategies when the system requirements and assumptions about the environment change. This framework is based on the notion of architectural homeostasis. 

Some earlier studies have shed some light on the existing work in the decentralization of self-adaptive systems. For instance,~\cite{Weyns2013-claims} performed a systematic literature review and found that 20\% of the studies that apply MAPE-based self-adaptation use multiple feedback loops. In~\cite{Weyns2013-patterns}, the authors present a simple notation for describing interacting MAPE loops and then use this language to specify a set of design patterns of interacting MAPE loops. Building upon this work,~\cite{ArcainiRS17} exploits abstract state machines to specify distributed and decentralized adaptation control in terms of MAPE-K control loops.  

Despite the variety of existing work, there is currently no clear overview of the state of the art in this area. Such an overview is important for researchers to grasp promising and recurring engineering practices to realize decentralized systems and drive future research efforts. 
To fill this gap, we conducted a mapping study~\cite{PETERSEN20151}. 
The goal of this mapping study is to structure the current knowledge on decentralization of self-adaptive systems (in contrast, a systematic review gathers and synthesizes evidence). Concretely, the study is centered on (i) motivations for choosing a decentralized approach to realize self-adaptation, (ii) the basic adaptation functions that are decentralized over different components, (iii) the realization of coordination between the components, and (iv) the open challenges in the area. 

The remainder of this paper is structured as follows. Section~\ref{sec:protocol} summarizes the protocol of the mapping study, including the research questions, the search strategy, and the data items we extracted. In Section~\ref{sec:results}, we present the results of the mapping study based on the extracted data. Section~\ref{sec:discussion} elaborates on the results and outlines a set of three coordination patterns for decentralized self-adaptive systems that emerged from the study. Finally, we wrap up with related work and conclusions in Sections~\ref{sec:related-work} and~\ref{sec:conclusions}. A reproduction package of the mapping study is available online~\cite{reproduction-package}.

\section{Summary Protocol}\label{sec:protocol}

We highlight the main parts of the study protocol. For the full protocol, we refer to the reproduction package~\cite{reproduction-package}.

\subsection{Scope of the Study}

In this mapping study we focus on architecture-based self-adaptive systems~\cite{garlan2004rainbow, Kramer2007SMS,weyns2012forms}, which is a widely applied approach to realize self-adaptation, see~\cite{weyns2020} for an overview. An architecture-based self-adaptive system consists of two essential parts: a \textit{managed system} that is controllable and subject to adaptation, and a \textit{managing system} that performs the adaptations of the managed system. The managed system operates in an \textit{environment} that is non-controllable. The managing system realizes four essential functions: Monitor-Analyze-Plan-Execute that share Knowledge~\cite{kephart2003vision}, MAPE in short. These components track the managed system and the environment, analyze the up-to-date knowledge and evaluate the options for adaptation based on the adaptation goals, generate a plan when needed, and execute this plan to adapt the managed system. MAPE provides a reference model that describes a managing system's essential functions and the interactions between them. A concrete architecture maps the functions to corresponding components\footnote{In the context of this study, a component implements one or multiple MAPE functions if these functions are centralized, or a part of one or multiple MAPE functions if the functions are decentralized.}, which can be a one-to-one mapping or any other mapping, such as a mapping of the analysis and planning functions to one integrated decision-making component. Our particular focus in this paper is on self-adaptive systems that realize one or more of the basic functions of self-adaptation (i.e., MAPE functions) by multiple components that coordinate with one another to realize that function. Examples include monitoring the managed system at distinct locations, analyzing adaptation options by multiple analysis components each having partial knowledge of the whole application, and making local adaptation decisions in different distributed components connected to parts of the application they manage.

\subsection{Goal and Research Questions}

We use the Goal-Question-Metric (GQM) approach~\cite{van2002goal} to formulate the goal of this mapping study: 

\begin{quote}
\textit{Purpose}: Organize and characterize \\
\textit{Issue}: the way decentralization of adaptation functions has been realized  \\
\textit{Object}: in research on self-adaptation \\
\textit{Viewpoint}: from a researcher’s viewpoint. 
\end{quote}

This overall goal translates in the following concrete research questions: 

\begin{enumerate}[label=\textbf{RQ\arabic*:}, leftmargin=*]
    \item What are the motivations to choose a decentralized approach for realizing self-adaptation?
    \item What basic adaptation functions of self-adaptive systems are decentralized over different components? 
    \item How is coordination managed between different components of a decentralized self-adaptive system?
    \item What are the open challenges reported for future work on decentralized self-adaptive systems?
\end{enumerate}

\subsection{Search Strategy}
To define the search query, we performed a pilot. We manually identified studies of decentralized self-adaptive systems published in 2019 and 2020 at SEAMS, TSE and TAAS. Then we defined an initial search query and used this to automatically search these venues. We gradually fine-tuned the query by comparing the results of the query to the manually identified studies until all papers of interest were returned by the automated query with minimal false positive results. The resulting search query is:\footnote{The query includes search terms related to distribution of self-adaptive systems. We made this decision consciously to ensure that all relevant studies are captured by the automated search, since some authors use the term distribution rather than decentralization to refer to self-adaptive systems where multiple components are used to realize single basic adaptation functions.} 
\vspace{5pt}\\
\texttt{\footnotesize((Title:decentral* OR Title:distrib*) AND \\\mbox{\ } (Title:adaptation OR Title:self*))}\\
\texttt{OR}
\texttt{\footnotesize((Abstract:decentral* OR Abstract:distrib*) AND \\\mbox{\ } (Abstract:adaptation OR Abstract:self*))}
\vspace{5pt}\\
To ensure primary studies of high quality and to manage the search process, we restricted the search to 25 well-known venues that publish studies on self-adaptation. The list includes venues on self-adaptation and autonomic computing (SEAMS, ACSOS, TAAS, etc.), software engineering (ICSE, ASE, FSE, TSE, TOSEM, etc.) software architecture (ECSA, ICSA, JSS, etc.), among others. For the full list, we refer to~\cite{reproduction-package}. 
In addition, we have applied snow-balling on all the papers that cited the community study on decentralized adaptation that emerged from the Dagstuhl seminar\footnote{\url{https://www.dagstuhl.de/en/program/calendar/semhp/?semnr=10431}} in 2010~\cite{Weyns2013-patterns}.

After collecting the primary studies with the search query, we apply a set of inclusion and exclusion criteria to each study. 

\subsection{Selection Criteria}

We apply the following inclusion criteria: 
\begin{itemize}
    \item[IC1] Papers written in English. 
    \item[IC2] Studies that apply architecture-based adaptation with MAPE-based feedback loops. 
    \item[IC3] Studies that decentralize at least one of the basic adaptation functions.
\end{itemize}

We apply the following exclusion criteria: 
\begin{itemize}
    \item[EC1] The paper is not a full research paper.
    \item[EC2] The paper presents a secondary study, e.g., literature review, or an overview of the field, e.g., a roadmap. 
\end{itemize}

A study was accepted if it satisfied all inclusion criteria and did not satisfy any of the exclusion criteria.

\subsection{Data Item to be Extracted}

To answer the research questions, we extracted the data items from the collected studies listed in Table~\ref{tab:data-items}. Only data explicitly stated in the studies was extracted.

\begin{table*}
    \centering
    \caption{Data items to be extracted.}
    \begin{tabular}{|c|p{4.5cm}|c|p{10.5cm}|}
    \hline
    \textbf{ID} & \textbf{Item} & \textbf{RQ} & \textbf{Explanation with Options} \\\hline
    D1 & Author(s) & N/A & Used for documentation purposes. \\\hline
    D2 & Publication year & N/A & Used for documentation purposes. \\\hline
    D3 & Title & N/A & Used for documentation purposes. \\\hline
    D4 & Venue & N/A & Used for documentation purposes. \\\hline
    D5 & Motivation for decentralization & RQ1 & The motivation for choosing a decentralized approach. Initial options are: scalability, reliability, and nature of the problem (e.g., inherent distribution of resources, multiple ownership). Further options are collected during data-extraction. \\\hline
    D6 & Trade-offs implied by decentralization & RQ1 & The trade-offs that are considered when choosing the possibility to apply decentralization. Options are collected during the data-extraction.\\\hline
    D7 & Basic adaptation functions decentralized & RQ2 & The specific MAPE functions that are realized in a decentralized manner. Options include: Monitoring, Analyzing, Planning, Executing, as well as combinations of these. \\\hline
    D8 & Interaction patterns applied & RQ2 & The patterns of~\cite{Weyns2013-patterns} used to decentralize basic adaptation functions. Options are: Coordinated Control, Information Sharing, Master/Slave, Regional Planning, Hierarchical Control. \\\hline
    D9 & Type of communication & RQ3 & The type of communication that components of decentralized adaptation functions use to interact. Initial options are: remote call, shared repository, explicit messages. Further options are collected during data-extraction.\\\hline
    D10 & Coordination mechanism & RQ3 & The mechanism used for coordination between components that realize basic adaptation functions. Initial options are: publish-subscribe, client-server, market mechanism. Further options are collected  during data-extraction.\\\hline
    D11 & Limitations & RQ4 & The reported limitations listed in the paper. Options are collected during the data-extraction.\\\hline
    D12 & Future work & RQ4 & Future work that is outlined in the paper. Options are collected during the data-extraction.\\\hline
    \end{tabular} \vspace{5pt}
    \label{tab:data-items}
\end{table*}

\section{Results of the Mapping Study}\label{sec:results}

\subsection{Demographic Information} 

Table~\ref{tab:number-of-papers} summarizes the selection of primary studies during the subsequent steps of the filtering process, based on the study protocol. In total, we extracted data of 96 primary studies.   

\begin{table}
    \centering
        \caption{Summary of the filtering steps.}
    \begin{tabular}{|p{6.8cm}|c|}
        \hline
        \textbf{Activity} & \textbf{\# Papers} \\\hline
        Automated search before filtering on venue & 32117 \\\hline
        Automated search after filtering on venue & 1211 \\\hline
        Paper selection via inclusion/exclusion criteria & 71 \\\hline
        Data extraction (deep text reading) & \emph{48} 
    \\\hline
        Snowballed papers of~\cite{Weyns2013-patterns} (1/2021) & 301 
        \\\hline
        Snowballed paper selection via inclusion/exclusion criteria & 186 \\\hline
        Data extraction snowballed papers (deep text reading) & \emph{48} \\\hline 
        
        Final number of selected primary studies & \emph{\underline{96}} \\\hline
    \end{tabular} \vspace{5pt}
    \label{tab:number-of-papers}
\end{table}

\begin{figure}
    \centering
    \includegraphics[width=1.0\linewidth]{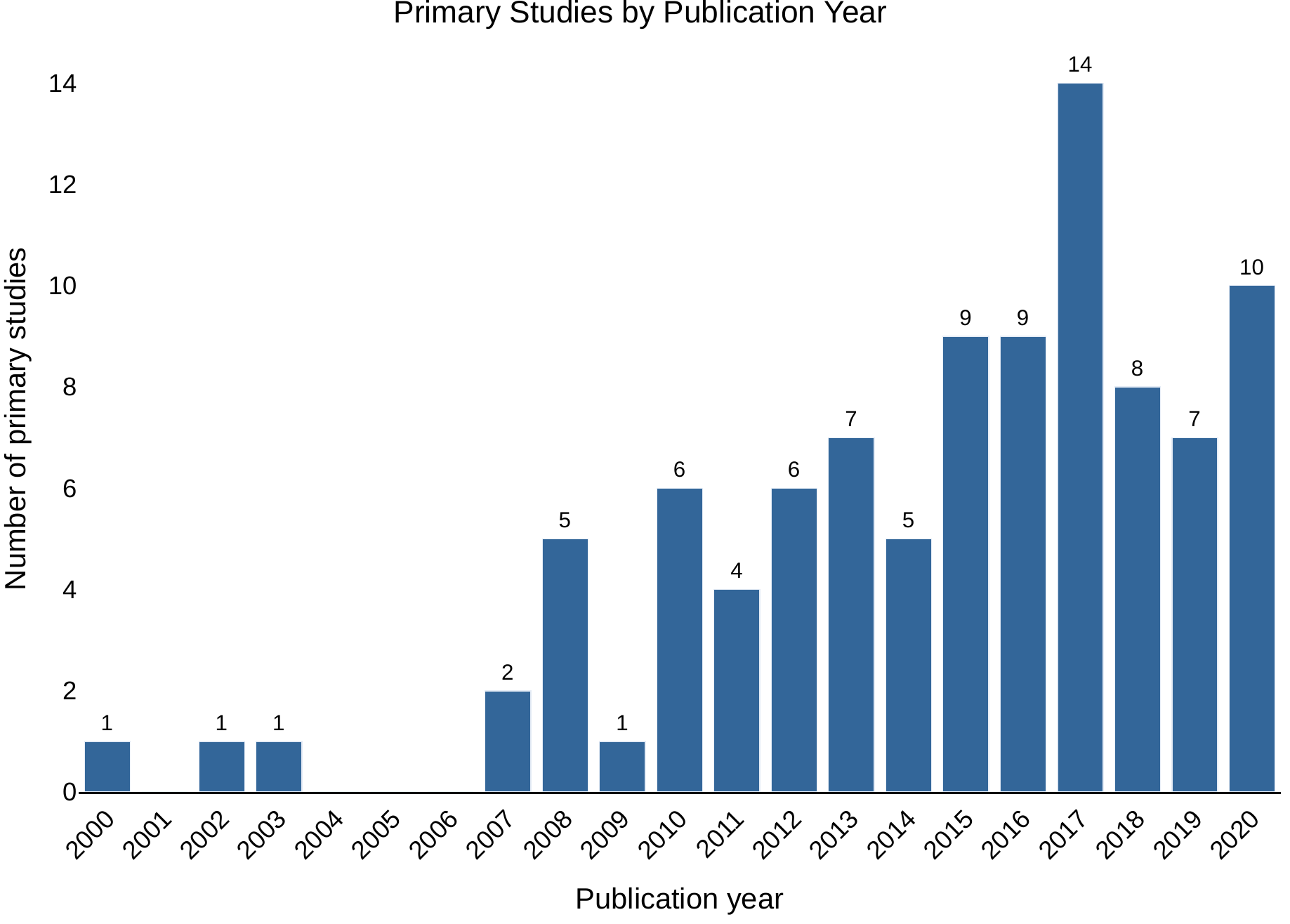}
    \caption{Overview of publication year of the selected primary studies.}
    \label{fig:publication-year}
\end{figure}

Figure~\ref{fig:publication-year} shows the distribution of papers over time. The graph shows a steady increase in the number of studies on decentralization of self-adaptation since 2007. The primary studies were published at 50 unique venues in total, showing the broad interest in research on decentralization of self-adaptation. The top venues are SEAMS (13 studies), ICAC and ACM TAAS (both seven studies), and SASO (five studies).

\subsection{What are the motivations to choose a decentralized approach for realizing self-adaptation?}

To answer the first research question, we collected data of the motivations for decentralizing self-adaptation functions (D5) and the tradeoffs implied by decentralization (D6). 

\begin{figure}
    \centering
    \includegraphics[width=0.75\linewidth]{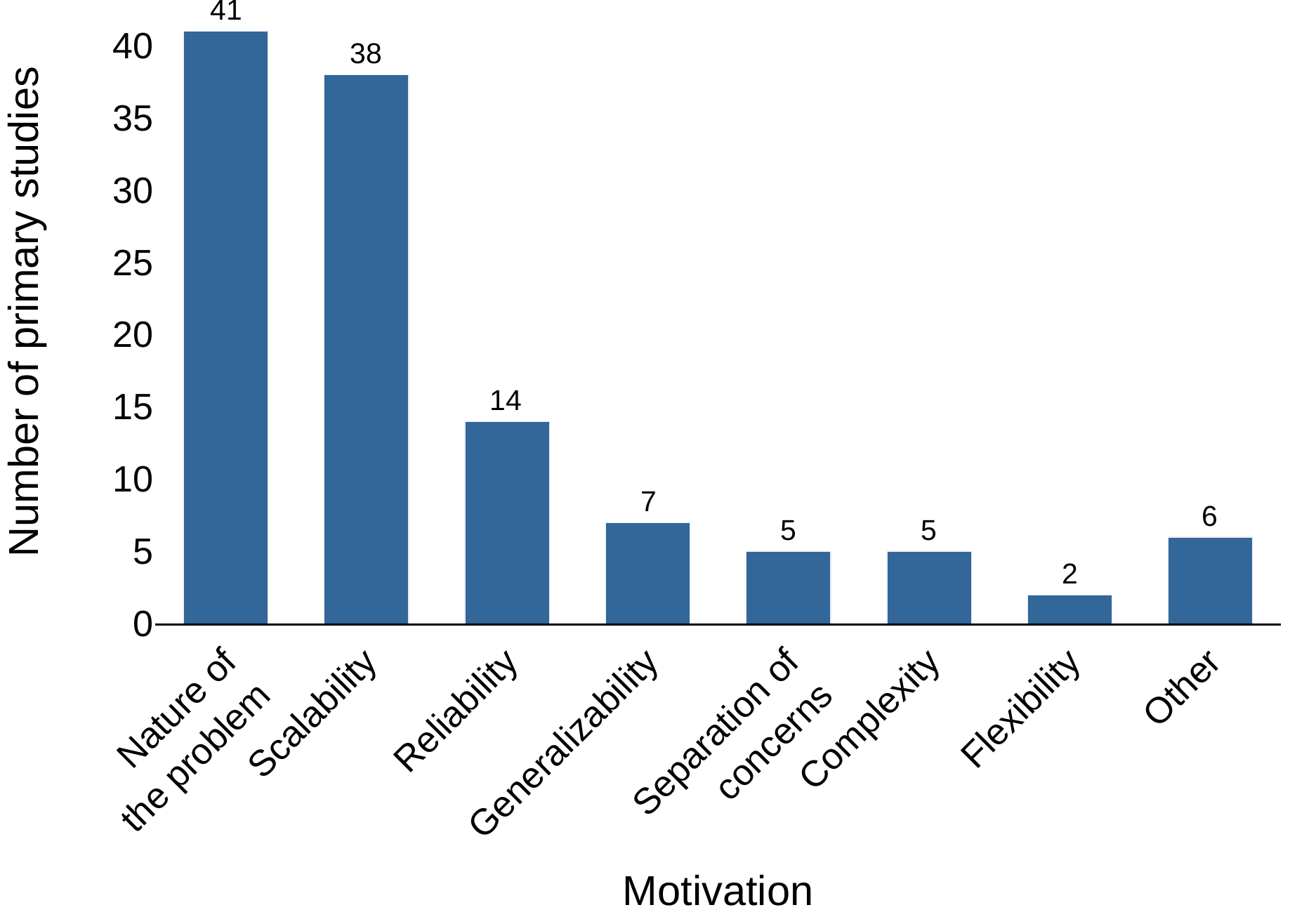}
    \caption{Reported motivations for using a decentralized solution.}
    \label{fig:motivations}
\end{figure}

Figure~\ref{fig:motivations} lists the reported motivations for using a decentralized solution to realize self-adaptation. 
The dominating motivation for decentralizing the self-adaptation logic is the nature of the problem (41 studies). For instance, in~\cite{Mejias2008}, the authors consider ad-hoc distributed networks where the topology and routing in the network is inherently organized on a peer to peer basis. 
In~\cite{Preisler2016}, the authors argue that modern 
computing systems are characterized by an increasing level of complexity and uncertain changes, requiring the need for a decentralized self-adaptation solution.
In~\cite{Gerostathopoulus2019}, the authors consider self-adaptation in a traffic setting of a city, where entities are inherently distributed and adaptation decisions are made by locally owned components.
Other top motivations are scalability (38 studies) and reliability (14 studies). Reliability often refers to the need to avoid a single point of failure. ``Other'' refers to other quality properties, such as interoperability, usability, and performance, accounting for six studies. 

Of the 96 primary studies, only 14 studies reported trade-offs implied by decentralization, see the summary in  Table~\ref{tab:trade-offs}. The dominating benefits of decentralization that are subject to trade-offs are scalability, and reliability, in particular robustness to a single point of failure. The main implications of decentralization that trade-off these benefits are related to the efficiency of the system, including overhead and suboptimality, design complexity, and potential security risks.

\begin{table}
    \centering
        \caption{Trade-offs implied by decentralizing adaptation functions.}
    \begin{tabular}{|c|p{3.2cm}|p{3.55cm}|}
    \hline
        \textbf{Study} & \textbf{Benefits decentralization} &  \textbf{Implications decentralization} \\\hline
        \cite{Cardei2000} & Reliability\newline No performance bottleneck\newline Flexibility & Performance cost adaptation\newline Latency to adapt/react \\\hline
        \cite{Sykes2011} & Scalability & Overhead     \\\hline
        \cite{Rekh2011} & Nature of problem & Potential security risks \\\hline
        \cite{Guinea2012} & Reliability\newline Reduced communication & Design complexity	 \\\hline
        \cite{Haupt2012} & Simple components \newline Scalability & Manageability\newline Design complexity   \\\hline
        \cite{Corina2012} & Reliability\newline No performance bottleneck & Consistency \\\hline
        \cite{Chen2014} & Scalability & Suboptimality \\\hline
        \cite{Kasinger2009} & Robustness\newline No performance bottleneck & Overhead \\\hline
        \cite{Mezghani2017} & Scalability & Suboptimality\newline Cost knowledge maintenance\\\hline
        \cite{Muccini2018} & Scalability & Cost of applying adaptation \\\hline
        \cite{Lesch2019} & Scalability & Suboptimality \\\hline
        \cite{DAngelo2017} & Generalizability & Design complexity \\\hline
        \cite{Rossi2020} & Scalability\newline Reliability & Consistency\newline Suboptimality\newline Design complexity  \\\hline
        \cite{Skandylas2020} & Scalability & Potential security risks \\\hline
    \end{tabular} \vspace{5pt}
    \label{tab:trade-offs}
\end{table}

We clarify trade-offs with two characteristic examples. In~\cite{DAngelo2017}, the authors put forward generalizability as a benefit for decentralization, more specifically, the benefits of collaboration in heterogeneous systems. However, decentralization introduces additional design complexity, i.e., developers need to answer questions such as: why should the system change, what changes in the system, how is change orchestrated, and which entities are responsible to perform the changes? 
On the other hand, in~\cite{Skandylas2020}, the motivation for decentralization is the scalability of the system. Yet, higher degrees of decentralization may imply security issues, in particular the average trust in the system may be affected by the topology of the decentralized system (and hence the communication paths).

\vspace{5pt}
\begin{tcolorbox}
\textbf{Answer to RQ1:} The main motivations to choose a decentralized approach for realizing self-adaptation are the nature of the problem (e.g., inherent distribution of services and resources, complexity of the system, multi-ownership), the scalability of the system, and its reliability. The main reported implications of decentralizing adaptation functions are cost in efficiency, increased design complexity, and security risks. 
\end{tcolorbox}

\subsection{What basic adaptation functions of self-adaptive systems are decentralized over different components?}

To answer the second research question, we collected data of the basic adaptation functions that were decentralized, i.e., realized by multiple components (D7), and the interaction patterns that were applied, based on~\cite{Weyns2013-patterns} (D8).

To characterize the decentralization of the basic adaptation functions, we distinguish between: 
(i) the number of instances that combinations of basic adaptation functions are jointly decentralized,\footnote{With an instance we mean a concrete application where adaptations functions are decentralized.} and (ii) the absolute number of occurrences that each distinct adaptation function has been decentralized for all instances reported in the primary studies.

\begin{figure}
    \centering
    \includegraphics[width=0.8\linewidth]{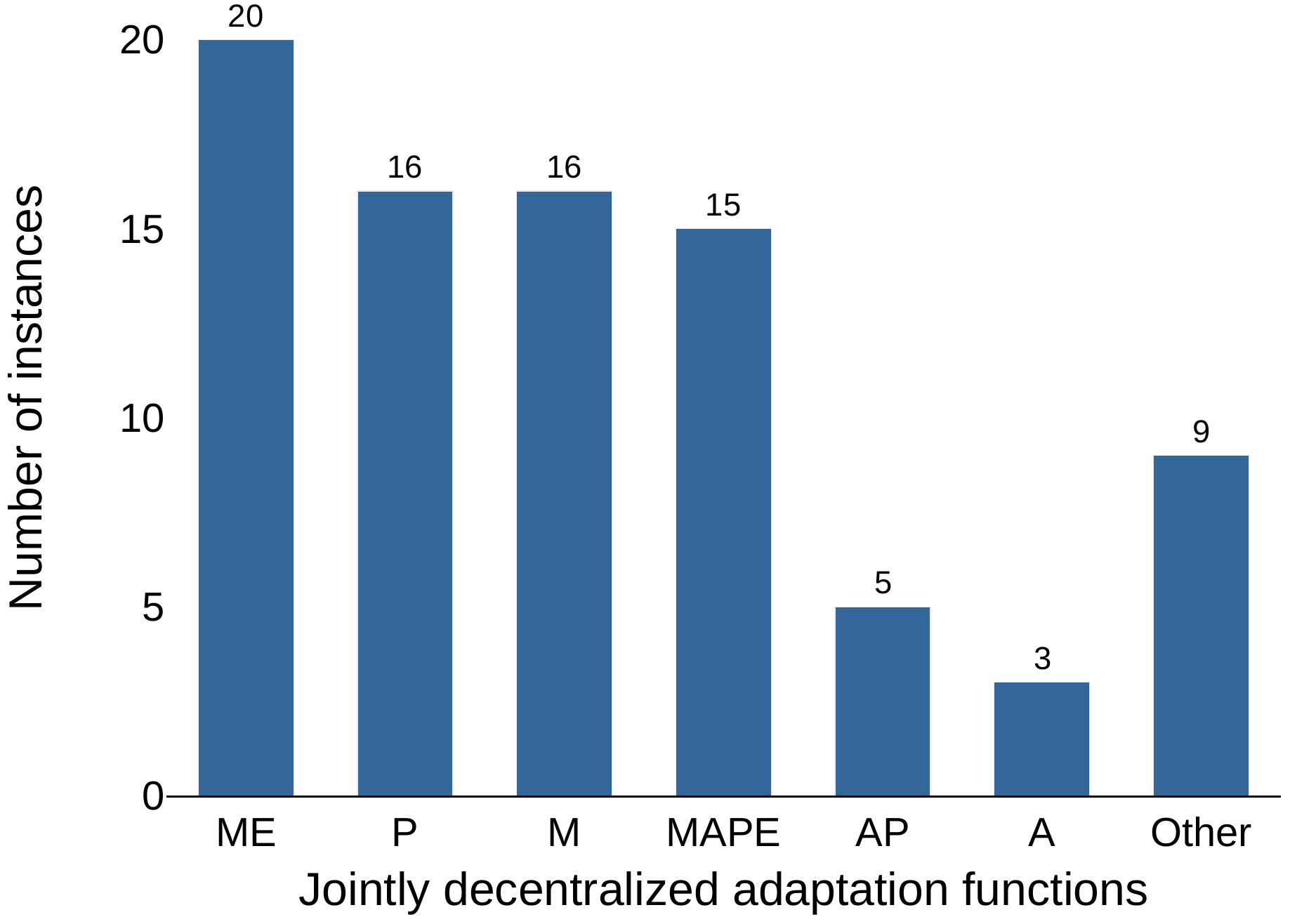}
    \caption{Jointly decentralized adaptation functions; entries are counted per instance of applications reported in the primary studies (total 102 instances).
    }
    \label{fig:cooperating-adaptation-functions}
\end{figure}

Figure~\ref{fig:cooperating-adaptation-functions} shows the frequencies that combinations of basic adaptation functions are jointly decentralized.
In total, 102 instances of concrete applications of decentralized adaptation functions were reported in the 96 primary studies.
The results show that most frequently the combination of the monitor and executor functions are decentralized (20 instances). The second most frequently applied decentralization is the combination of all MAPE functions of the feedback loop (15 instances). The combination of analysis and planning was applied in 5 instances. Besides combinations of basic adaptation functions, decentralization was also applied to only monitoring and planning both in 16 instances, and analysis in 3 instances. The category ``Other'' groups the remaining instances that applied different combinations of decentralized adaptation functions, but each combination was applied in at most two instances.

\begin{figure}
    \centering
    \includegraphics[width=0.65\linewidth]{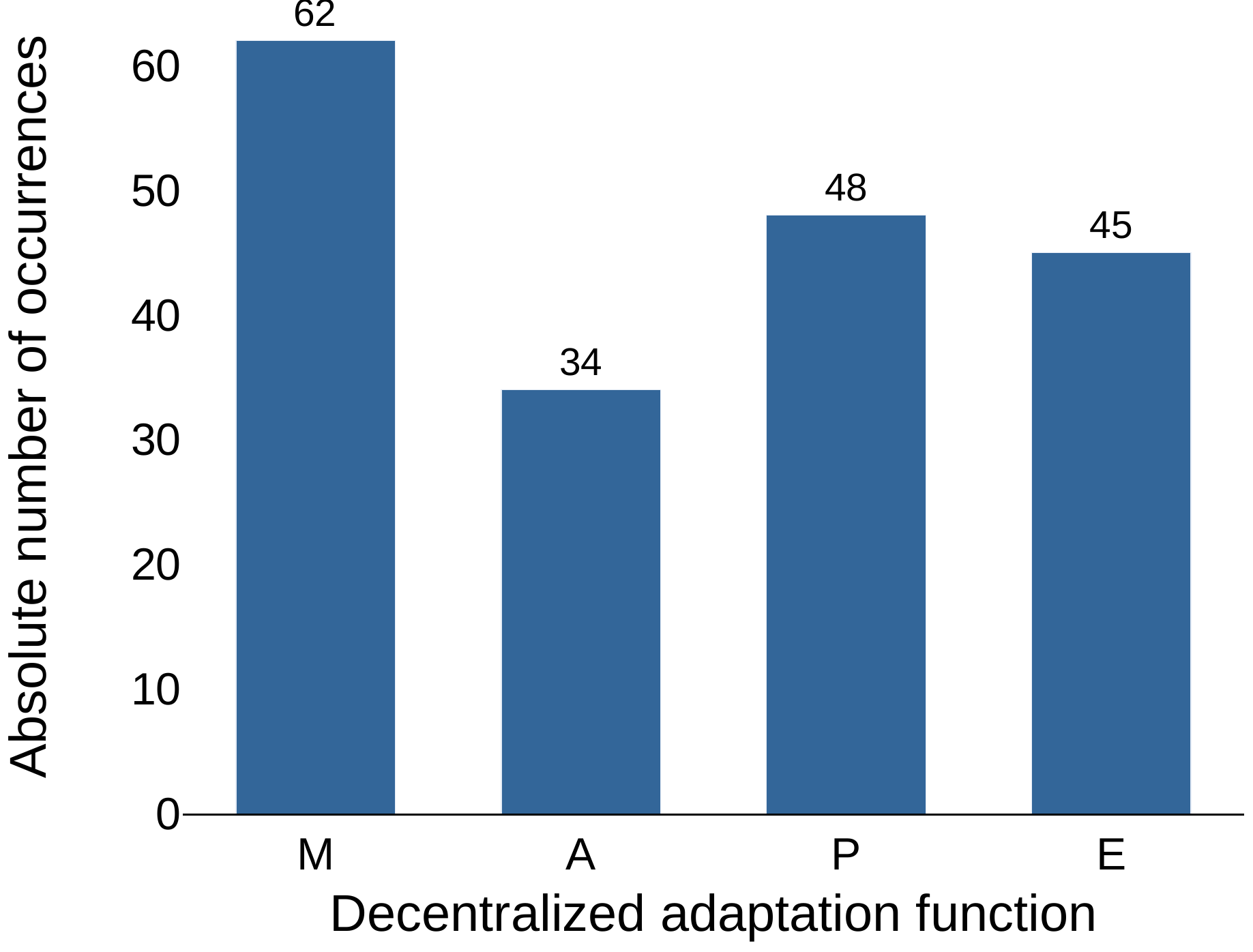}
    \caption{Distinct adaptation functions that are decentralized; entries are counted per occurrence in instances of applications reported in the primary studies (total 189 occurrences of 102 instances).}
    \label{fig:single-adaptation-functions}
\end{figure}

Figure~\ref{fig:single-adaptation-functions} shows that all four basic functions are frequently decentralized. 
In total, we counted 189 occurrences of distinct adaptation functions that were decentralized for the 102 instances. Monitoring with 62 occurrences and planning with 48 occurrences have been decentralized mostly. 

Besides the decentralization of concrete adaptation functions, we also looked at generic solutions that support the instantiation of different decentralization schemes that combine adaptation functions. We found in total 18 such solutions. 
A typical example, described in~\cite{ArcainiRS17}, offers a modeling framework for decentralized self-adaptive systems that can be instantiated for different combinations of MAPE functions that are decentralized. Another example is~\cite{Watzoldt2015} that extends the Eurema~\cite{Vogel2014} framework with support for interactions between feedback loops to achieve coordinated self-adaptation.

\begin{figure}
    \centering
    \includegraphics[width=0.7\linewidth]{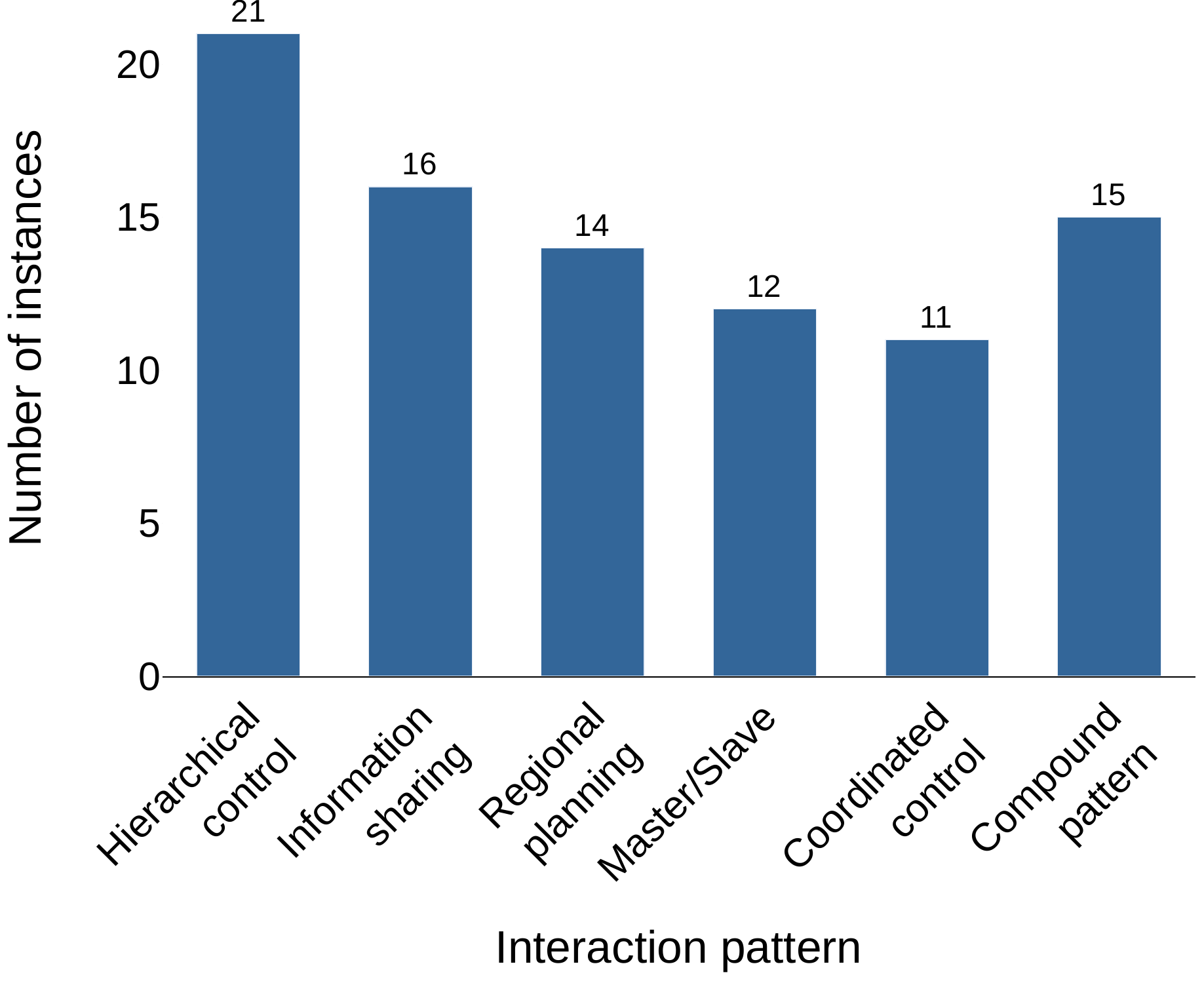}
    \caption{Applied interaction patterns in the primary studies (total 93 instances).}
    \label{fig:interaction-patterns}
\end{figure}

Figure~\ref{fig:interaction-patterns} shows the frequencies of interaction patterns of~\cite{Weyns2013-patterns} that are applied in instances of decentralized adaptation of the primary studies. Remarkable, in total, 74 of the 102 instances applied one of the patterns of~\cite{Weyns2013-patterns}. All five patterns are regularly applied, with hierarchical control the most frequently one (21 instances). In addition, 15 instances combined two patterns of~\cite{Weyns2013-patterns}. 
For instance,~\cite{Ismail2015} combines information sharing and regional planning in a multi-cloud setting. Each cloud is equipped with a set of local adaptation managers that manage different types of resources (virtual machines, platforms, etc.). These adaptation managers apply the information sharing pattern to exchange information within the cloud. In addition, the adaptation managers can coordinate with peers on other clouds to generate adaptation plans for their adaptation tasks. 
In~\cite{Hachicha2018} the authors propose a compound pattern that composes the master/slave pattern and the coordinated control pattern. The aim of this hybrid approach is to enhance the master component in the master-slave pattern to exchange information with analyzer and planning components of the coordinated control pattern. This way, the master component can acquire global information to make more optimal adaptation decisions.

The remaining 28 instances that have not used any of the patterns of~\cite{Weyns2013-patterns} applied different combinations of decentralized adaptation functions.  
For instance, in~\cite{Yamagata2020} the authors decentralize the combination of the monitor and analysis functions.

\vspace{5pt}
\begin{tcolorbox}
\textbf{Answer to RQ2:} All adaptation functions of self-adaptive systems are decentralized, with monitoring and planning being the most popular. The main jointly decentralized adaptation functions are monitoring combined with executing, all MAPE functions, and analysis combined with planning. Monitoring and planning are frequently decentralized alone. The majority of the primary studies apply one of the patterns of~\cite{Weyns2013-patterns}, or they combine two of this set of patterns. 
\end{tcolorbox}

\subsection{How is coordination managed between different components of a decentralized self-adaptive system?}

To answer the third research question, we collected data of the type of communication that MAPE components use to interact (D9) and the mechanism used for coordination (D9). 

\begin{figure}
    \centering
    \includegraphics[width=0.7\linewidth]{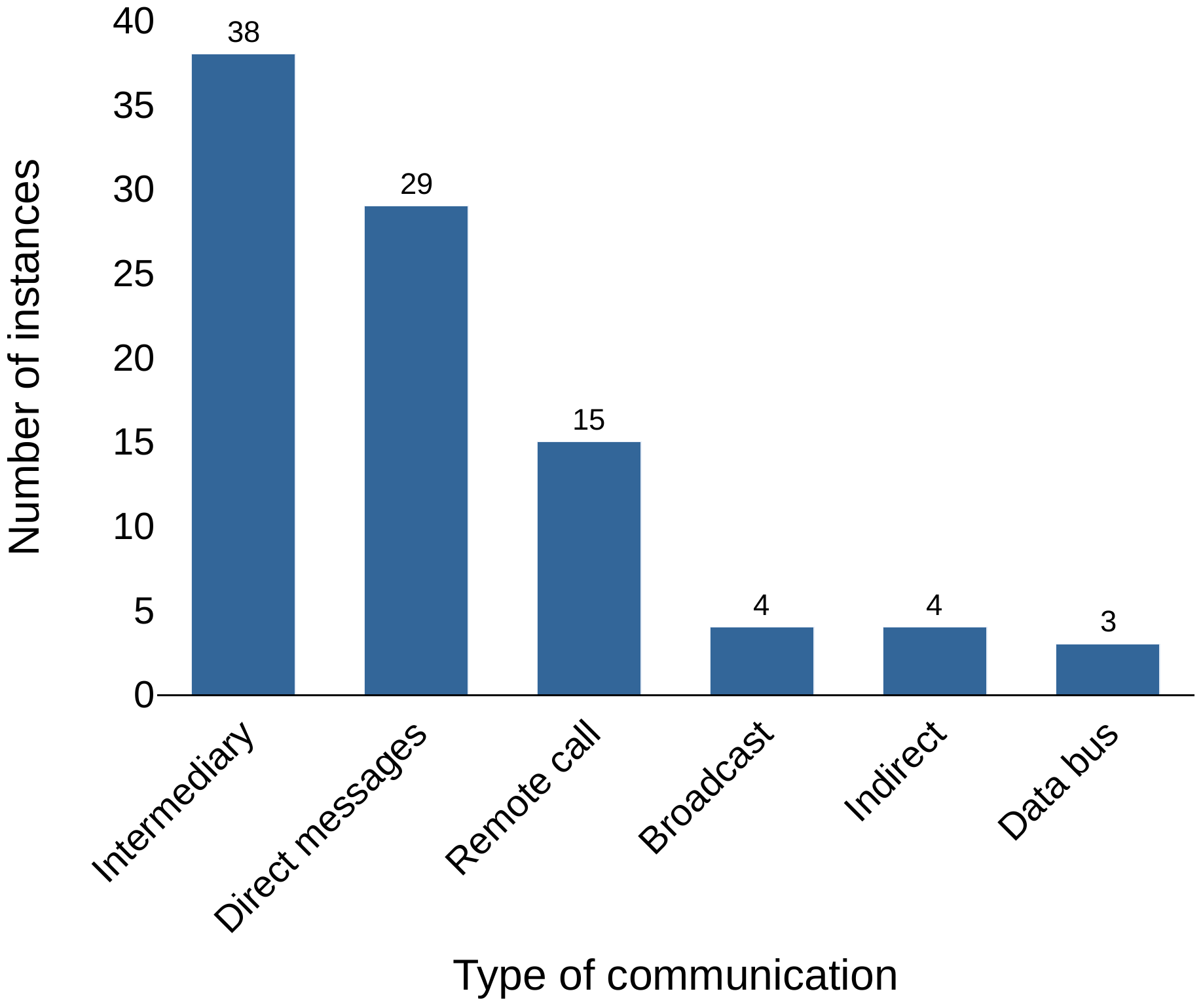}
    \caption{Types of communication applied in the primary studies (93 instances).}
    \label{fig:type-of-communications}
\end{figure}

Figure~\ref{fig:type-of-communications} shows the different types of communications that are used by components of decentralized MAPE functions to interact. The main type of communication is through a messaging system that acts as an intermediary (38 instances). For instance,~\cite{Baresi2011} uses a middleware for communication, more concretely REDS (Reconfigurable Dispatching System), that implements a distributed publish-subscribe mechanism.
Direct communication is also often used to 
by components of decentralized MAPE functions to interact (29 instances). For instance, in~\cite{Jahan2020} a coordination control component composes a list of adaptation options and sends this list directly to two feedback loops that deal with different concerns. The feedback loops score the adaptation options according to their own goals and send the results to the coordination control component that then decides on the adaptation decision.

Other types of communication that are used are remote call (15 instances) and broadcast (4 instances). In four instances, the interaction between components of decentralized MAPE functions happened indirectly, examples are~\cite{AlShistawy2008, Cesari2014}. In these cases no explicit forms of communication are used to convey information, but rather a form of indirect communication is used were the effects of adaptation decisions by one particular feedback loop are observed by other feedback loop(s).

\begin{figure}
    \centering
    \includegraphics[width=1\linewidth]{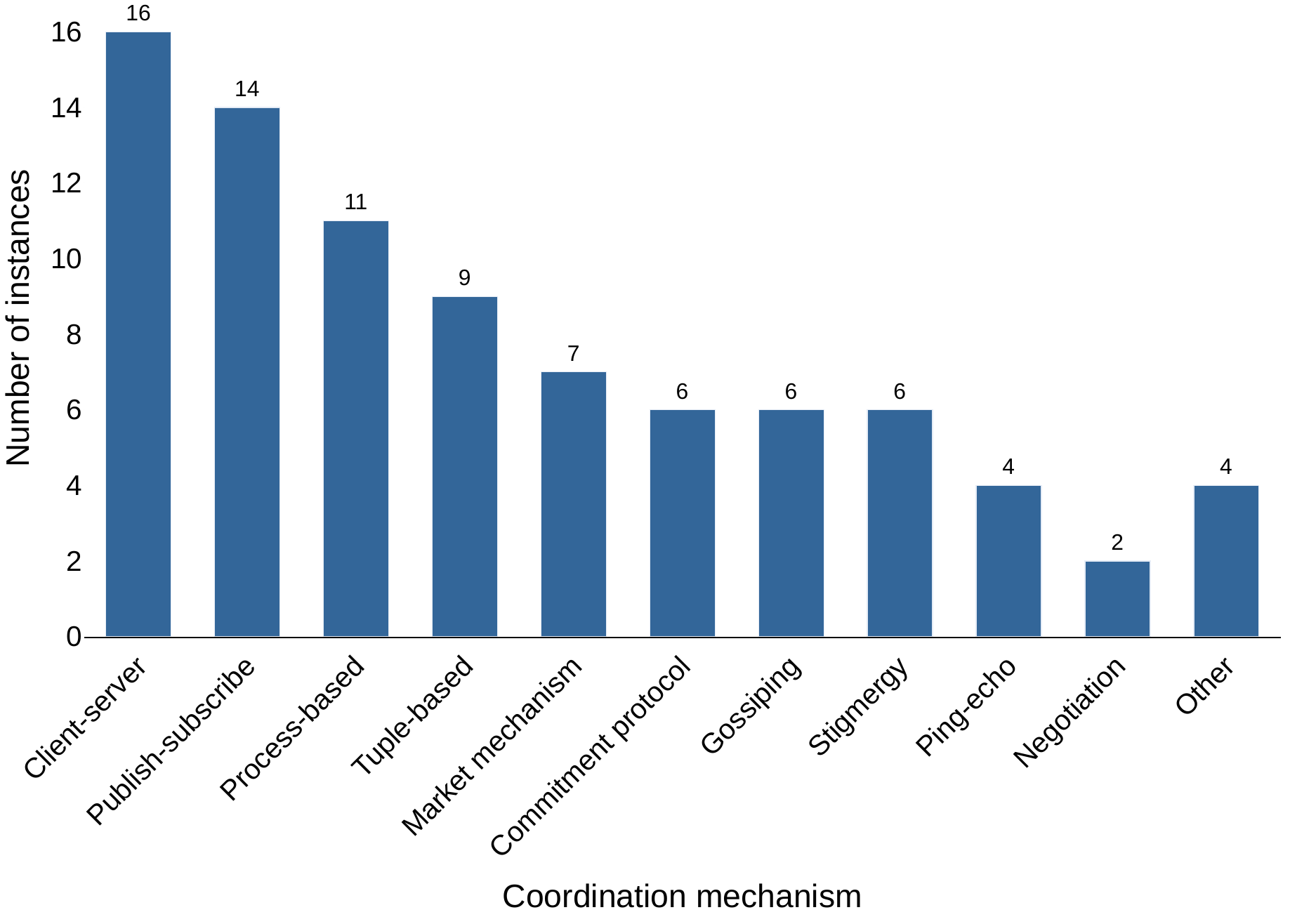}
    \caption{Coordination mechanisms applied in the studies (85 instances).}
    \label{fig:coordination-mechanisms}
\end{figure}

Figure~\ref{fig:coordination-mechanisms} shows the coordination mechanisms applied by components of decentralized self-adaptive system. The results show that a wide variety of coordination mechanisms are used. Most popular is client-server (16 instances) followed by publish-subscribe (14 instances). Process-based coordination\footnote{Process-based coordination relies on clearly defined interfaces, stream or channel connections between producers and consumers~\cite{PAPADOPOULOS1998329}} was used in 11 instances. For example, in~\cite{Hachicha2018} the authors combine master-slave and coordinated control of~\cite{Weyns2013-patterns} in one compound interaction pattern where MAPE components expose and interact through well-defined interfaces. 
Among the other coordination mechanisms used, we highlight Stigmergy with 6 instances. For instance,~\cite{Viroli2016} uses aggregate computing principles to realize self-adaptation. Through the specification of collective behavior for each MAPE component individually, multiple components for each adaptation function are instantiated. The collective behavior in the system then results in self-adaptation that emerges.

\vspace{5pt}
\begin{tcolorbox}
\textbf{Answer to RQ3:} Components of decentralized self-adaptive systems use a broad set of coordination mechanisms to work together and realize the basic adaptation functions. Most frequently used are client-server, publish-subscribe, and process-based coordination. These coordination mechanisms rely on different types of communication, with intermediary, direct messages, and remote call as the most frequently used. 
\end{tcolorbox}

\subsection{What are the open challenges reported for future work on decentralized self-adaptive systems?}

To answer the fourth research question, we collected data of reported limitations (D11) and suggestions for future work reported in the primary studies (D12).

Of the 96 primary studies, only 26 reported limitations of the proposed work. Note that we focus here on limitations with respect to the decentralization of self-adaptation only. Table~\ref{tab:limitations} lists the types of limitations described in the studies. 
The main reported limitation is modeling restrictions with 10 occurrences. For instance,~\cite{ArcainiRS17} states that the proposed framework for decentralized self-adaptation does not support modeling of uncertainty and timing aspects of the decentralized system. 
The second most reported limitation is underlying assumptions with 6 occurrences. For instance, the solution presented in~\cite{Gothel2017} assumes that communication works perfectly and components do not fail. Overhead and inefficiencies is another limitation. 

\begin{table}[b!]
    \centering
        \caption{Reported types of limitations.}
    \begin{tabular}{|p{4cm}|c|}
        \hline
        \textbf{Type of Limitation} & \textbf{\# Occurrences} \\\hline
        Modeling restrictions  & 10 \\\hline
        Underlying assumptions  & 6 \\\hline        
        Overhead and inefficiencies & 5 \\\hline
        Impact on guarantees & 3 \\\hline
        Conflicting decisions & 2 \\\hline
    \end{tabular}
    \label{tab:limitations}
\end{table}

Of the 96 primary studies, 
58 outline suggestions for future work on the topic of decentralization of self-adaptation. Table~\ref{tab:futurework} lists the main topics for future research in the area.

The main topic put forward for future research is to enhance the mechanisms used for coordination between the components that realize the adaptation functions. For instance,~\cite{Nallur2013} proposes to extend mechanisms for trading between components by considering reputation. Another example is~\cite{Capra2002} that proposes to extend market mechanism with policy strategies to deal with conflicts. Besides coordination between components, authors also argue for further research on the decision-making process by the components themselves. For instance,~\cite{Nardelli2019} argues for exploring more advanced decentralized policies with different (possible conflicting) optimization objectives. The two highest ranked topics for future research may be related to the reported limitations of modeling restrictions and underlying assumptions. 
Providing guarantees is another important topic for future research, which directly relates to the limitation of impact on guarantees. For instance,~\cite{Anders2011} argues for using a model checker to ensure behavioral guarantees of coalition formation. Other high ranked topics for future research are improving overall system goals, in particular to deal with trade-offs between goals and deal with new goals such as robustness (this topic relates to the limitation overhead and inefficiencies), and dealing with conflicts, in particular with respect to decisions and the adaptation actions implied by the decisions (which relates to the limitation conflicting decisions). Remarkably only two studies highlight the need for a disciplined engineering approach to realize decentralized self-adaptive systems~\cite{Yamagata2020,Weyns2008}, and  only two studies emphasize the need to investigate emergent behavior~\cite{Watzoldt2015,Li2018}.   

\begin{table}[t!]
    \centering
        \caption{Suggested topics for future research.}
    \begin{tabular}{|p{5cm}|c|}
        \hline
        \textbf{Topic for Future Research} & \textbf{\# Occurrences} \\\hline
        Enhance coordination mechanisms  & 14 \\\hline
        Improve decision-making process  & 12 \\\hline      
        Improve on overall system goals & 12 \\\hline
        Provide guarantees & 11 \\\hline
        Deal with conflicting decisions  & 8 \\\hline
        Deal with overhead & 6 \\\hline
        Deal with uncertainties & 4 \\\hline
        Provide disciplined engineering approach & 2 \\\hline
        Investigate emergent behavior & 2 \\\hline
    \end{tabular}
    \label{tab:futurework}
\end{table}

\vspace{5pt}
\begin{tcolorbox}
\textbf{Answer to RQ4:} 
The key open challenges for decentralized self-adaptive systems are to enhance coordination mechanisms, improve the decision-making process, improve on the overall system goals, and provide guarantees for the system goals. Only a few studies highlight the need for a disciplined engineering approach to realize decentralized self-adaptive systems and the same applies to investigate emergent behavior.  
\end{tcolorbox}

\section{Discussion}\label{sec:discussion}

\subsection{From Interaction Patterns to Coordination Patterns}\label{subsec:patterns} 

The patterns specified in the community paper~\cite{Weyns2013-patterns} only specify abstract interactions between MAPE components that are realized in a decentralized way. Hence, we refer to these as ``interaction patterns.'' Since a majority of the primary studies apply one of these patterns and given that we have collected also data about the way these abstract interactions are realized we investigated whether we can identify any emerging patterns in the way the MAPE components of particular interaction patterns coordinate to realize decentralization. We refer to these as ``coordination patterns''. 

Table~\ref{tab:interaction-coordination} shows a mapping between the interaction patterns and coordination mechanisms. We highlight three specific mappings that have occurred multiple times: Information sharing \& Tuple-based coordination, Regional planning \& Market mechanism, Hierarchical control \& Client server. In this paper, we provide high-level descriptions of the patterns, a full fledged specification of the patterns is left as future work. 

\begin{table*}
    \centering
    \caption{Observed mapping of interaction patterns and coordination mechanisms.}
    \begin{tabular}{|c|c|c|c|c|c|c|c|c|c|}
        \cline{2-10}
        \multicolumn{1}{c|}{} & 
        \begin{tabular}{@{}c@{}}\textbf{Client}\\\textbf{server}\end{tabular} &
        \begin{tabular}{@{}c@{}}\textbf{Publish}\\\textbf{subscribe}\end{tabular} &
        \begin{tabular}{@{}c@{}}\textbf{Process}\\\textbf{based}\end{tabular} & 
        \begin{tabular}{@{}c@{}}\textbf{Tuple}\\\textbf{based}\end{tabular} & 
        \begin{tabular}{@{}c@{}}\textbf{Market}\\\textbf{mechanism}\end{tabular} &
        \begin{tabular}{@{}c@{}}\textbf{Commitment}\\ \textbf{protocol}\end{tabular} &
        \textbf{Gossiping} & 
        \textbf{Stigmergy} &  
        \textbf{Negotiation} \\\hline
        
        \textbf{Coordinated control} & 1 & 0 & 2 & 1 & 2 & 1 & 2 & 1 & 1 \\\hline
        \textbf{Information sharing} & 0 & 2 & 3 & \cellcolor{gray!20}5 & 1 & 0 & 2 & 1 & 1 \\\hline
        \textbf{Regional planning} & 2 & 2 & 1 & 1 & \cellcolor{gray!20}4 & 1 & 0 & 0 & 1 \\\hline
        \textbf{Master-Slave} & 3 & 3 & 1 & 1 & 0 & 1 & 1 & 0 & 0 \\\hline
        \textbf{Hierarchical control} & \cellcolor{gray!20}6 & 4 & 2 & 2 & 0 & 4 & 0 & 1 & 0 \\\hline
    \end{tabular}
    \label{tab:interaction-coordination}
\end{table*}

\subsubsection{Information Sharing Coordination Pattern}

\begin{figure}
    \centering
    \includegraphics[width=1\linewidth]{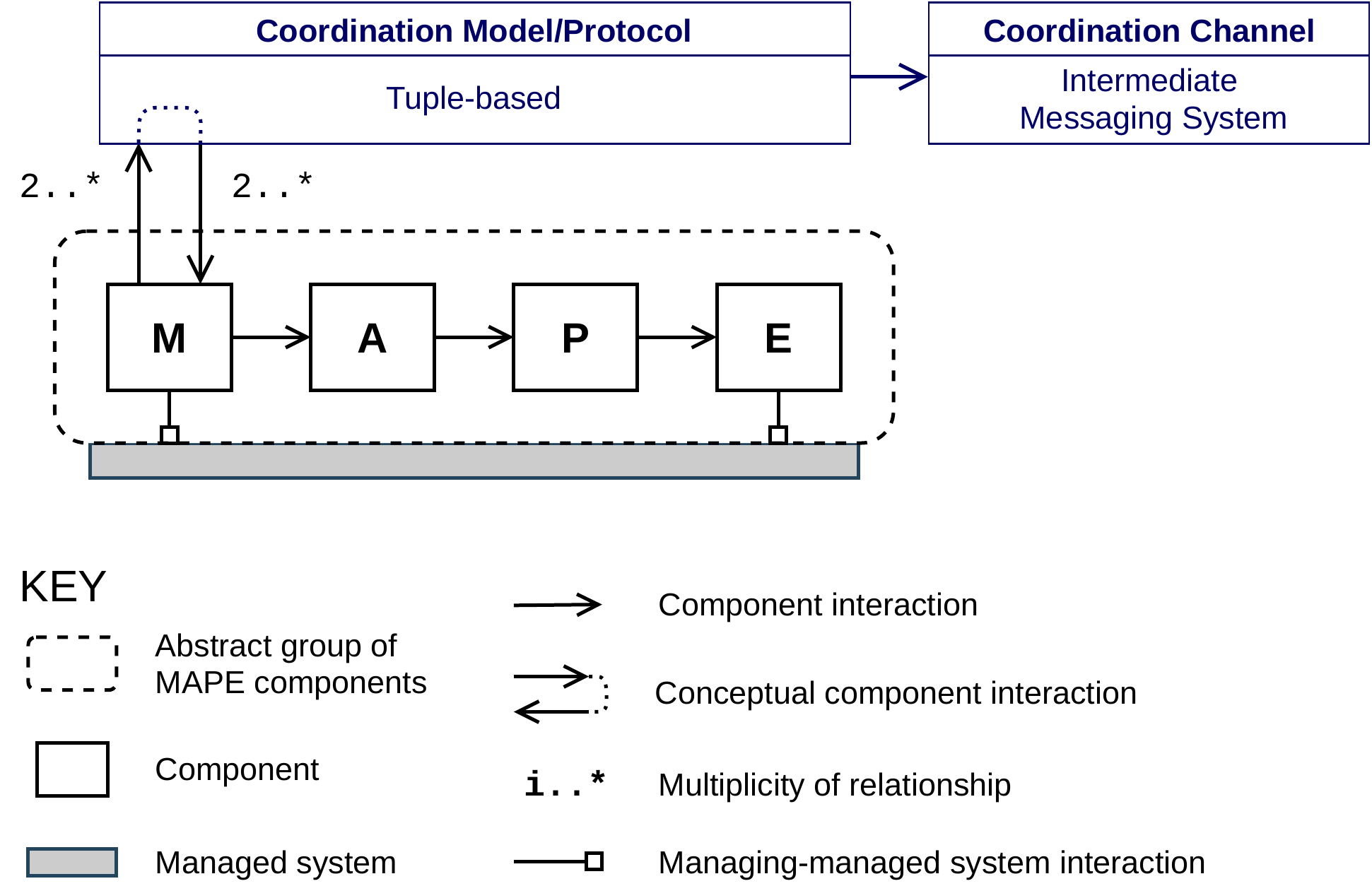}
    \caption{Information sharing coordination pattern.}
    \label{fig:information_sharing}
\end{figure}

Figure~\ref{fig:information_sharing} illustrates the information sharing coordination pattern that combines the  information sharing interaction pattern with tuple based coordination. We combine here the notation introduced in~\cite{Weyns2013-patterns} with the representation of a coordination mechanism described in the  FORMS reference model~\cite{weyns2012forms}, in particular the unification of the reflection and distribution perspective with the MAPE-K perspective.
Information sharing realizes the monitor function by multiple  components that can exchange information to realize adaptation of a managed system that is typically distributed in a network. The pattern supports scalability and fast adaptation response, but decisions may not always be optimal if only partial information is available locally. 
In the information sharing coordination pattern, multiple monitor components coordinate using a tuple based coordination model with an associated protocol (e.g., put tuple, copy tuple, take tuple, etc.) using an intermediate messaging system (such as a distributed tuple space).

As an example, in~\cite{Florio2016} the authors present Gru, an approach to deal with decentralized adaptation of a micro-service architecture. The system consists of nodes that each run a Gru agent. Each node registers itself in a repository. During monitoring, nodes can request discovery information about other nodes in the system in order to exchange data directly.

\subsubsection{Regional Planning Coordination Pattern}

\begin{figure}
    \centering
    \includegraphics[width=0.9\linewidth]{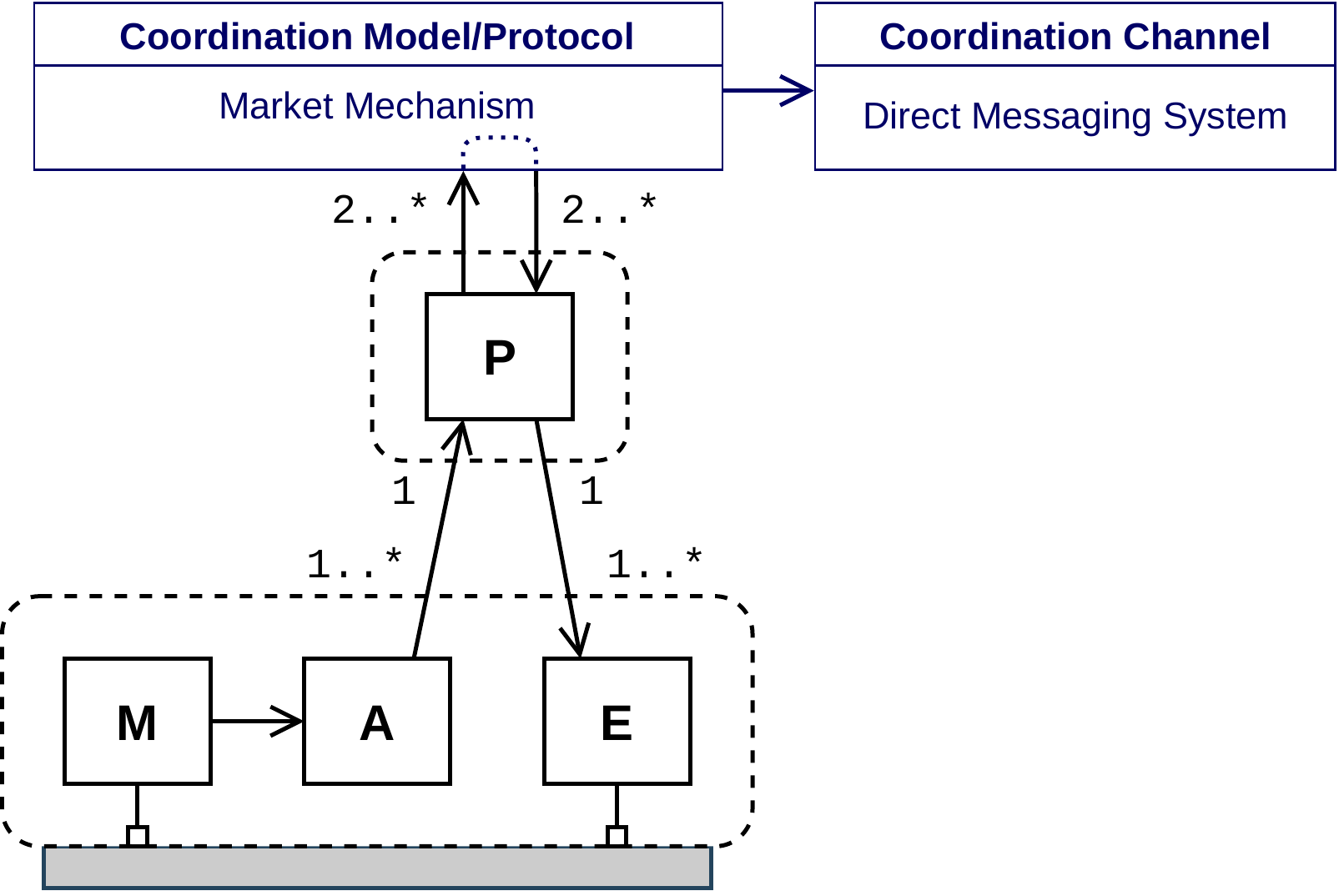}
    \caption{Regional planning coordination pattern.}
    \label{fig:regional_planning}
\end{figure}

Figure~\ref{fig:regional_planning} shows the regional planning  coordination pattern that combines the regional planner interaction pattern with coordination using a marked mechanism.
Regional planning supports the planning of local adaptations for loosely coupled parts of a managed system (regions), taking into account effects that cross the different parts (between regions). An example is a multi-cloud setting where owners of regions may want to work together without exposing too much information.
In the regional planning coordination pattern, multiple planner components coordinate using a market mechanism based on a particular protocol (e.g., send order, send bid, assign winner, etc.) using a direct messaging system that enables the planners to communicate through the exchange of messages.

As an example,~\cite{Nallur2013} describes adaptation within a multi-cloud system. Each cloud that corresponds to a region specifies requirements for its applications in terms of computing and storage resources. To achieve these requirements at runtime in the face of uncertainties, the multi-cloud system orchestrates self-adaptation through multiple types of agents that coordinate using a market mechanism. Application Agents, local to each cloud, monitor the applications and check whether the Quality of Service and cost requirements are met or could be optimized. If the requirements are not met, a Buyer Agent is invoked that acts on behalf of the cloud (region). If redundant services are available, a Seller Agent is informed that can sell services for a specified price on behalf of the cloud. Market Agents are responsible for handling trading rounds in the market between Buyer Agents and Seller Agents. The services sold vary in terms of cost and Quality of Service. Buyer Agents analyze the services and rank bids according to the offered Quality of Service and cost, finally choosing a new service that can be deployed on a cloud to run the application.

\subsubsection{Hierarchical Control Coordination Pattern}

\begin{figure}
    \centering
    \includegraphics[width=1\linewidth]{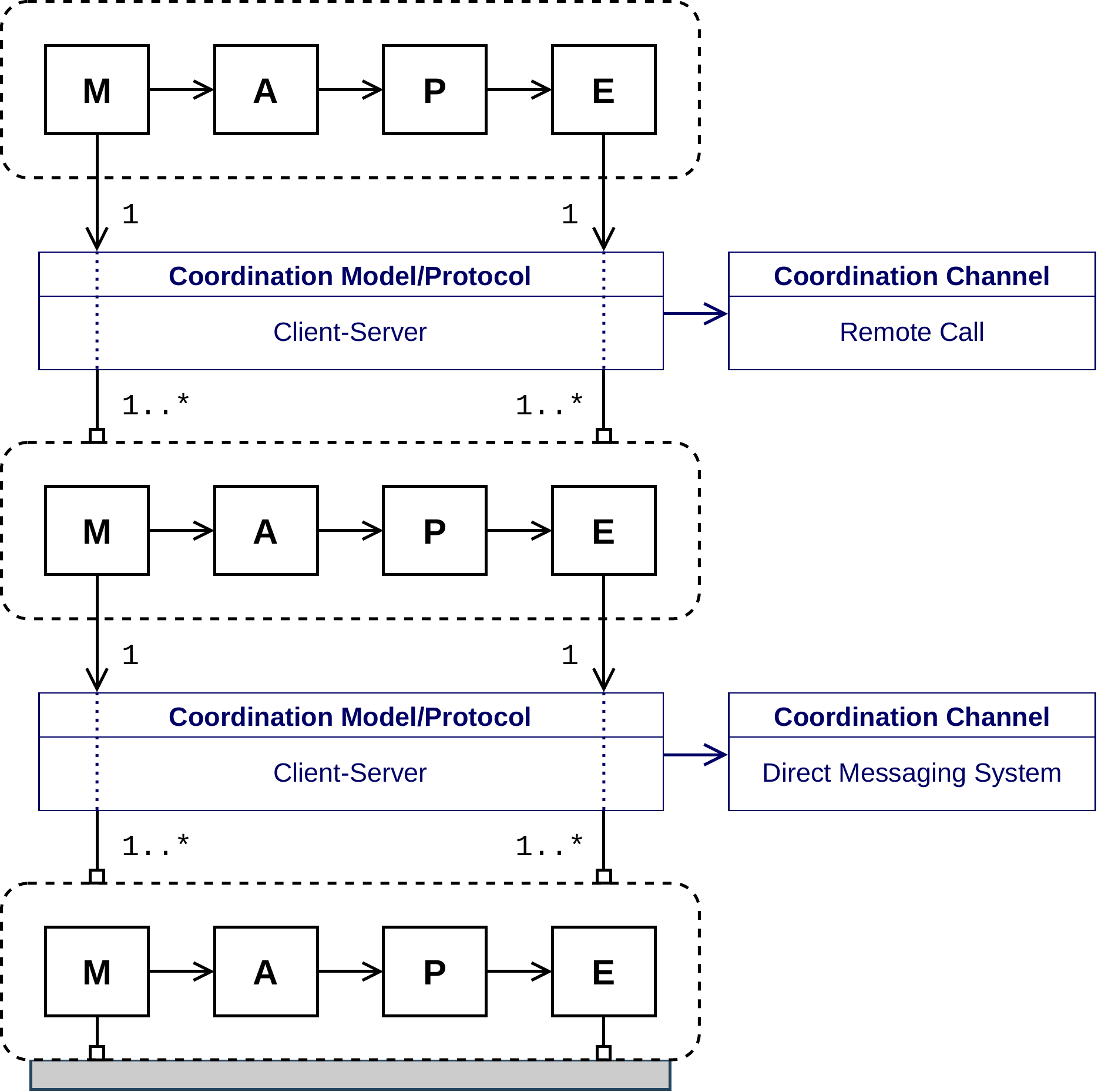}
    \caption{Hierarchical control coordination pattern.}
    \label{fig:hierarchical_control}
\end{figure}

Figure~\ref{fig:hierarchical_control} shows the hierarchical control coordination pattern (here illustrated for three layers) that combines the hierarchical control interaction pattern with client-server coordination. 
The hierarchical control coordination pattern provides a layered separation of concerns to manage the complexity of self-adaptation. The pattern structures the adaptation logic as a hierarchy of MAPE-based loops that typically focus on different concerns at different levels of abstraction, and may operate at different time scales. In the hierarchical control coordination pattern, the monitor and execute components at the bottom layer directly interact with the managed subsystem, while the components of higher-level layers interact with feedback loops at the layers beneath.
The layers coordinate using a client server coordination mechanism, where layer $N$ serves as a client that makes use of layer $N-1$ that serves as a server when realizing the overall functionality of adaptation. The coordination channel can be organized using remote calls or a direct messaging system. 

As an example,~\cite{Rossi2020} presents a two-layered hierarchy that controls the scaling of replicas per micro-service in a micro-service architecture. The feedback loop at the top, denoted as the Application Manager, provides input for the feedback loops at the bottom, denoted as Micro-service Managers. The loop at the top does not necessarily run as frequently as the loops at the bottom. Concretely, the Application Manager monitors the performance of the application, analyzes this information and constructs a global policy for the Micro-service Managers. This global policy steers the overall system adaptation by providing guidelines to the Micro-service Managers.

\subsection{Threats to Validity}\label{subsec:validity} 

To ensure quality and soundness of the results of this mapping study,
we followed a systematic approach documented in a protocol. However, there are some possible threats to validity.

\textit{Internal validity} refers to the extent to which a causal conclusion based on a study is warranted. This mapping study targets decentralization of self-adaptation for systems that rely on architecture-based adaptation. Identifying whether a study follows this paradigm is sometimes not easy as papers may leave this information implicit in the text. To address this threat, the decisions to include or exclude papers with respect to the paradigm used was made by multiple researchers of the team involved in this study. In particular, for the first 20 papers, two researchers made individual decisions about including/excluding the papers. Then the results were discussed and differences were resolved. The results were then crosschecked by the third researcher involved in the study. This approach was repeated for an additional set of 10 papers. The results show that we then reached consensus on including/excluding papers. Of the remaining batch of papers, another six papers where discussed before including/excluding them to ensure consistent results. 

\textit{External validity} refers to the extent to which the findings can be generalized to all architecture-based self-adaptation research. By limiting the automatic search to selected venues and applying an automatic search strategy using a selection of search engines, we may have missed some primary studies. To mitigate this threat, we started the search process with pilot searches to define and tune the search string. In addition, we performed snowballing of the community paper on patterns for decentralized adaptation to find potentially missed studies.

\textit{Construct validity} refers to the extent to which we obtained the right measure and whether we defined the right scope in relation to what is considered research on decentralized self-adaptation.  Regarding the scope of architecture-based, we acknowledge that other types of adaptation exist, such as control-based adaptation and self-organisation. Our choice to limit the scope to architecture-based adaptation is motivated by the wide spread usage of the approach on the one hand, and the difference in nature of other related paradigms on the other hand. There may also be a validity threat regarding the quality of reporting of studies that may have affected both the selection of papers and the extraction of data. To anticipate this threat, we extracted data from well-known venues that publish research on self-adaptation.  As for the papers found through snowballing, we applied a cursory quality check based on the quality criteria for reporting studies used for instance in~\cite{DYBA2008833,6224395,TSE.2017.2704579} (based on a score for the reported problem definition, problem context, research design, contributions, insights, limitations). This way, we excluded two papers from the mapping study. Overall, we found that the reporting quality of the included primary studies was good, providing a basis to make sound conclusions about the validity of extracted data.

\textit{Reliability} refers to the extent to which we can ensure that our results are the same if our study would be conducted again. The researchers that performed this mapping study may have been biased when collecting and analyzing data of primary studies. To address this threat, we defined a protocol that we followed carefully throughout the study. Nevertheless, the background and experience of the researchers may have created some bias. In addition, we have made all the material of the study available for other researchers.

\section{Related work}\label{sec:related-work}

To the best of our knowledge, there is no secondary study on the decentralization of MAPE-based self-adaptive systems. Hence, we focus on a selection of related studies that elaborate on why and how feedback loops in self-adaptation can be decentralized and what the problems are associated with it. 
		
Both in~\cite{Andersson2009} and~\cite{Brun2009}, the authors argue that the degree of decentralization is a key dimension of  self-adaptive systems. In~\cite{Andersson2009}, the authors approach decentralization from a modeling perspective, emphasizing the mechanisms that are used to react to change, while~\cite{Brun2009} approaches decentralization from the design of feedback loops emphasising the potential of  natural systems to achieve fault tolerance and scalability. These basic works offer a context in which our mapping study is situated and highlight the relevance of providing a systematic overview of the state of the art of decentralized self-adaptive systems.

A pioneering work that aimed at consolidating knowledge on decentralization of self-adaptive systems is~\cite{Weyns2013-patterns}. This paper introduces a 
systematic approach for describing different levels of decentralized control in self-adaptive systems and use that to document a number of existing patterns
of interacting MAPE-based feedback loops. 
The authors of~\cite{Waetzoldt:2014pa} take a complementary angle and classify runtime models that capture the shared knowledge employed by different components of feedback loops that make decentralized adaptation decisions. Our mapping study aligns with these efforts, yet it provides a systematic map of the work on decentralized self-adaptation. 

In~\cite{2489850.2489860}, the authors present three architectural styles to realize self-adaptation
in systems of systems, i.e., systems that integrate independently useful systems into a larger system. These styles are ``Local Adaptations,'' ``Regional Monitoring–Local Adaptations,'' and ``Collaborative Adaptations.'' These three styles provide increasing levels of knowledge sharing and collaboration, allowing to mitigate uncertainty at different scales. In the local adaptations style, feedback loops do not interact with each other. This maximizes the autonomy to designers of individual systems, but the cost is a tradeoff with respect to providing guarantees about uncertainties of cross-system properties. In the collaborations style, feedback loops interact directly with each other and collaboratively realize adaptations. This style creates dependencies between individual systems, but offers designers better support to guarantee cross-system properties. In the regional monitoring style, feedback loops only share information, providing a middle ground between the other two styles. Our mapping study complements this work with an in-depth study of decentralization of MAPE-based feedback loops.

\section{Conclusions and Future Work}\label{sec:conclusions}

The goal of this paper was to draw a landscape of the state of the art in decentralized self-adaptive systems. To that end, we performed a mapping study.  
The results show a steady increase in research on decentralization of self-adaptation published across a broad range of venues. 

Decentralization of self-adaptation is primarily motivated by the nature of the problem, the need to deal with scale and ensure reliable solutions. Yet decentralization may imply a cost in efficiency (both in introduced overhead and suboptimal adaptation), an increase of design complexity, and potential security risks. All adaptation functions of self-adaptive systems are frequently decentralized. The main jointly decentralized functions are monitoring combined with executing, all MAPE functions, and analysis combined with planning. Components of decentralized self-adaptive systems use a broad set of coordination mechanisms to work together, with client-server, publish-subscribe, and process-based coordination being the most frequently used. The key reported open challenges for decentralized self-adaptive systems are enhancing coordination mechanisms, improving the decision-making process, and providing guarantees for the systems goals. 
Other potential interesting open challenges that need to be tackled are dealing with emergent behavior, and study and apply learning and search-based techniques in decentralized settings. 

The data extracted from this mapping study allowed us to enhance the patters specified in~\cite{Weyns2013-patterns}, a widely cited paper that resulted from a community seminar at Dagstuhl. In particular, we extended a subset of the original interaction patterns with coordination mechanisms. Concretely, we identified three ``coordination patterns'' that have been applied multiple times in primary studies: information sharing coordination, regional planning coordination, and hierarchical control coordination.

We offer the insights obtained from this mapping study together with the open challenges as starting points for future research in the exciting area of decentralized self-adaptation.

\bibliographystyle{abbrv}
\bibliography{main}
\end{document}